\documentclass[final,3p,times,twocolumn]{elsarticle}

\usepackage{graphicx}
\usepackage{amsmath, amssymb}
\usepackage{mathrsfs}
\usepackage{bm}
\usepackage{braket}
\usepackage{multirow}
\usepackage[usenames,dvipsnames]{color}
\usepackage[colorlinks=true, citecolor=blue]{hyperref}
\usepackage{listings}

\biboptions{sort&compress}

\newcommand{\diff}{\mathrm{d}}

\newcommand{\real}{\mathrm{Re}\,}

\newcommand{\imu}{\mathrm{i}}
\newcommand{\epn}{\mathrm{e}}

\newcommand{\ua}{\uparrow}
\newcommand{\da}{\downarrow}
\newcommand{\dg}{\dagger}
\newcommand{\la}{\langle}
\newcommand{\ra}{\rangle}

\newcommand{\sg}{\sigma}

\newcommand{\gm}{\gamma}

\newcommand{\T}{\mathrm{T}}
\newcommand{\dvec}[1]{\hspace{-1mm}\stackrel{\leftrightarrow}{#1}\hspace{-1mm}}
\newcommand{\nt}{\notag \\}

\newcommand{\mrm}[1]{\mathrm{#1}}
\newcommand{\mcal}[1]{\mathcal{#1}}

\lstdefinelanguage{julia}
{
  keywordsprefix=\@,
  morekeywords={
    calc_density, write_density, calc_rgrid, calc_wannier_matrix, write_wannier_matrix,
    using
  },
  sensitive=true,
  morecomment=[l]{\#},
  morestring=[b]',
  morestring=[b]" 
}

\newcounter{bla}

\journal{Computer Physics Communications}

\begin{document}

\begin{frontmatter}



\title{DiracBilinears.jl: A package for computing Dirac bilinears in solids
}


\author[a]{Tatsuya Miki\corref{author}}
\author[b]{Hsiao-Yi Chen}
\author[c]{Takashi Koretsune}
\author[b]{Yusuke Nomura}

\cortext[author] {Corresponding author.\\\textit{E-mail address:} t.miki.315@ms.saitama-u.ac.jp}
\address[a]{Department of Physics, Saitama University, Sakura, Saitama 338-8570, Japan}
\address[b]{Institute for Materials Research, Tohoku University, Sendai, Miyagi 980-8577, Japan}
\address[c]{Department of Physics, Tohoku University, Sendai, Miyagi 980-8578, Japan}

\begin{abstract}

DiracBilinears.jl is a Julia package for computing Dirac bilinears, which are fundamental physical quantities of electrons in relativistic quantum theory, using first-principles calculations for solids.
In relativistic quantum theory, 16 independent bilinears can be defined using the four-component Dirac field.
We take the non-relativistic limit for the bilinears, which corresponds to the $1/m$ expansion, and focus on the low-energy physics typically considered in condensed matter physics.
This package can evaluate the spatial distributions and Wannier matrix elements of the Dirac bilinears in solids quantitatively by connecting to the external first-principles calculation packages, including Quantum ESPRESSO, Wannier90, and wan2respack.

% A submitted program is expected to satisfy the following criteria: it must be of benefit to other physicists, or be an exemplar of good programming practice, or illustrate new or novel programming techniques which are of importance to computational physics community; it should be implemented in a language and executable on hardware that is widely available and well documented; it should meet accepted standards for scientific programming; it should be adequately documented and, where appropriate, supplied with a separate User Manual, which together with the manuscript should make clear the structure, functionality, installation, and operation of the program.

% Your manuscript and figure sources should be submitted through Editorial Manager (EM) by using the online submission tool at \\
% https://www.editorialmanager.com/comphy/.

% In addition to the manuscript you must supply: the program source code; a README file giving the names and a brief description of the files/directory structure that make up the package and clear instructions on the installation and execution of the program; sample input and output data for at least one comprehensive test run; and, where appropriate, a user manual.

% A compressed archive program file or files, containing these items, should be uploaded at the "Attach Files" stage of the EM submission.

% For files larger than 1Gb, if difficulties are encountered during upload the author should contact the Technical Editor at cpc.mendeley@gmail.com.

\end{abstract}


\end{frontmatter}

% \linenumbers

% All CPiP articles must contain the following
% PROGRAM SUMMARY.

{\bf PROGRAM SUMMARY}
% NEW VERSION PROGRAM SUMMARY}
  %Delete as appropriate.

\begin{small}
\noindent
{\em Program Title:} DiracBilinears.jl                                          \\
{\em CPC Library link to program files:} (to be added by Technical Editor) \\
{\em Developer's repository link:} \url{https://github.com/TatsuyaMiki/DiracBilinears.jl.git}
% (if available) 
\\
{\em Code Ocean capsule:} (to be added by Technical Editor)\\
{\em Licensing provisions:} GNU General Public License 3.0 \\
% CC0 1.0/CC By 4.0/MIT/Apache-2.0/BSD 3-clause/BSD 2-clause/GPLv3/GPLv2/LGPL/CC BY NC 3.0/MPL-2.0  \\
{\em Programming language:} Julia                                   \\
{\em External software:} \texttt{Quantum ESPRESSO}, \texttt{Wannier90}, \texttt{wan2respack} \\
% {\em Supplementary material:}                                 \\
  % Fill in if necessary, otherwise leave out.
% {\em Journal reference of previous version:}*                  \\
  %Only required for a New Version summary, otherwise leave out.
% {\em Does the new version supersede the previous version?:}*   \\
  %Only required for a New Version summary, otherwise leave out.
% {\em Reasons for the new version:*}\\
  %Only required for a New Version summary, otherwise leave out.
% {\em Summary of revisions:}*\\
  %Only required for a New Version summary, otherwise leave out.
{\em Nature of problem:
% (approx. 50-250 words):
}
In relativistic quantum theory, Dirac bilinears are the fundamental physical quantities derived from the Dirac field.
This package is a tool for the evaluation of the bilinears in solids quantitatively.
\\
  %Describe the nature of the problem here. \\
{\em Solution method:
% (approx. 50-250 words):
}
This package evaluates Dirac bilinears in solids by applying non-relativistic limit expressions, focusing on the low-energy regime typically discussed in condensed matter physics.
It uses results from first-principles calculations performed with \texttt{Quantum ESPRESSO}, \texttt{Wannier90}, and \texttt{wan2respack}.
By using the Bloch wave functions and the Wannier functions obtained from these packages, this package computes spatial distributions and Wannier matrix elements of the bilinears.
\\
  %Describe the method solution here.
{\em Additional comments including restrictions and unusual features:
% (approx. 50-250 words):
}
This package requires \texttt{Quantum ESPRESSO} calculations using norm-conserving pseudopotentials and supports \texttt{wan2respack} in ``spinor'' branch on GitHub \footnote{GitHub repository of the \texttt{wan2respack} ``spinor'' branch: \url{https://github.com/respack-dev/wan2respack/tree/spinor}.}.
  %Provide any additional comments here.
   \\

\end{small}

\section{Introduction \label{sec:intro}}

The asymmetry in electronic states underlies a variety of phenomena in solid-state systems and serves as a crucial element for understanding material behavior.
Recently, materials with polar, chiral \cite{Naaman15, Tokura18, Naaman20, Tokura21, Bloom24, Gohler11, Brinkman24, Sakano20, Inui20, Rao19}, and ferro-axial \cite{Yamagishi23, Zhang15, Hayashida20, Hayashida21, Hayashida23, Jin20, Sayantika24, Hlinka16} structures have gathered significant attention due to their novel electronic properties and potential functionalities.
The behaviors of these asymmetric materials are often described by fundamental physical quantities, such as charge, spin, and current density, which can exhibit non-trivial spatial distributions.

In previous studies by one of the authors and his collaborators, the microscopic physical quantities derived from the Dirac field in relativistic quantum theory have been investigated \cite{Hoshino23, Hoshino24, Miki24}.
Relativistic quantum theory enables the definition of not only charge, spin, and current density but also electron chirality and polarity.
These quantities are part of the Dirac bilinears, which are classified based on their symmetry properties under spatial inversion, time reversal, and whether they are scalar or vector \cite{Berestetskii_book, Sakurai_book}.
In particular, the electron chirality characterizes the electronic states of chiral and axial materials, as demonstrated through the first-principles calculations.
Additionally, the relation between electron chirality and circular dichroism in photoemission spectroscopy, a phenomenon unique to chiral materials, was discussed \cite{Miki24}.

In the context of quantum chemistry, it has been pointed out that the electron chirality at nuclei is related to the parity-violating energy difference between enantiomers \cite{Senami19, Kuroda22}.
It has also been discussed that the spatial gradient of the electron chirality density induces an additional spin torque \cite{Tachibana12, Hara12, Fukuda16}.
Moreover, the spatial distribution of the electron chirality in molecules has been investigated \cite{Fukuda16} using \texttt{QEDalpha} package \cite{QEDalpha}, based on the wave functions obtained from the quantum chemistry calculation package \texttt{DIRAC} \cite{Dirac13}.

\begin{table*}[tb]
    \centering
    \begin{minipage}{0.4\textwidth}
        \begin{tabular}{c|c|c}
            {\bf Scalars} & NRL & SI/TR \\ \hline
           $\mathcal S,\ \rho/e$ & $\psi^\dg\psi$ & $+/+$ \\
           $\mcal P$ & $-\frac{\hbar}{2mc} \nabla\cdot (\psi^\dg \bm \sg \psi)$ & $-/-$ \\
           $\mcal \tau^Z$ & $\frac{1}{2mc}\psi^\dg \dvec{\bm p} \cdot \bm \sg \psi$ & $-/+$ \\
        \end{tabular}
    \end{minipage}
    \hspace{-40pt}
    \begin{minipage}{0.4\textwidth}
        \begin{tabular}{c|c|c}
            \begin{tabular}{c}
                {\bf Three-component} \\ 
                {\bf vectors}
            \end{tabular} & NRL & SI/TR \\ \hline
           $\bm j/ce$ & $\frac{1}{2mc}\psi^\dg \dvec{\bm p}\psi$ & $-/-$ \\
           $\bm{\mcal A},\ \bm{\mcal M}$ & $-\psi^\dg \bm \sg \psi$ & $+/-$ \\
           $\bm{\mcal P}$ & $\frac{\hbar}{2mc}\nabla(\psi^\dg \psi) + \bm{\mcal P}_S$ & $-/+$ 
        \end{tabular}
    \end{minipage}
    \caption{Scalars and vectors in Dirac bilinears and their parity under spatial inversion (SI) / time reversal (TR).
    As a reference, we also list the expressions of non-relativistic limit (NRL).}
    \label{tab:bilinears}
\end{table*}

In this paper, we present DiracBilinears.jl, a Julia package for calculating the Dirac bilinears derived from the Dirac field.
We apply the non-relativistic limit to focus on low-energy physics.
This package can evaluate the spatial distributions of the bilinears and their Wannier matrix elements by connecting the open-source packages for the first principles calculations \texttt{Quantum ESPRESSO} \cite{Giannozzi09}, \texttt{Wannier90} \cite{Mostofi08, Pizzi20}, and \texttt{wan2respack} \cite{Kurita23}.
It is worth noting that similar calculations for spatial distributions can also be performed using the \texttt{FLPQ} module~\cite{FLPQ} in \texttt{QEDalpha} package~\cite{QEDalpha}, which has been developed independently.
The \texttt{QEDalpha} package serves as a post-processing tool for the open-source first-principles calculation package \texttt{OpenMX}~\cite{OpenMX, Ozaki03, Ozaki04}, which employs localized basis sets.

This paper is organized as follows.
In the next section, we explain the theoretical background related to the microscopic physical quantities, which can be calculated using DiracBilinears.jl.
Section~\ref{sec:usage} provides a detailed explanation of the installation procedure and usage of this package. 
In Sec.~\ref{sec:application}, we present several examples as applications of the package. 
Finally, we summarize this paper in Sec.~\ref{sec:summary}.

\section{Theoretical background \label{sec:theory}}

In this section, we explain the theoretical background.
First, we provide definitions of Dirac bilinears and their non-relativistic limits which can be calculated using DiracBilinears.jl (Sec.~\ref{sec:bilinears} and Sec.~\ref{sec:nrl}).
Then, we describe how to evaluate the spatial distribution of the bilinears in Sec.~\ref{sec:bloch}.
We also provide explanation of the Wannier matrix elements in Sec.~\ref{sec:wannier}.

\subsection{Dirac bilinears \label{sec:bilinears}}

Dirac bilinears are fundamental quantities in relativistic quantum field theory, constructed from the Dirac field $\Psi$.
For comparison, let us first recall the bilinears for the two-component Schr\"odinger field of electrons in the non-relativistic case.
In the non-relativistic framework, these bilinears are classified using the Pauli matrices $\sg^i$ with $i = 1, 2, 3$ (or equivalently $x, y, z$).
Specifically, there are $2\times 2 = 4$ independent bilinears: (three-component) spin density $\psi^\dg \bm \sg \psi$ and charge density $\psi^\dg \psi$.

In the relativistic case, on the other hand, the Dirac field $\Psi$ is described as a four-component spinor that satisfies the Dirac equation.
The bilinears for the Dirac field are constructed using the gamma matrices, defined as
$
\gm^0 = 
\begin{pmatrix}
    1 & 0 \\
    0 & -1
\end{pmatrix}, 
\gm^i = 
\begin{pmatrix}
    0 & \sg^i \\
    -\sg^i & 0
\end{pmatrix}$
, and $
\gm^5 = -\imu\gm^0 \gm^1 \gm^2 \gm^3 =
\begin{pmatrix}
    0 & -1 \\
    -1 & 0
\end{pmatrix}$.
Using these gamma matrices $\gamma^\mu\ (\mu=0,1,2,3)$ and $\gamma^5$, we can construct $4 \times 4 = 16$ independent bilinears, which are referred to as Dirac bilinears \cite{Berestetskii_book, Hoshino24}: 
\begin{align}
    \mcal S &= \bar \Psi \Psi
    , \label{eq:db_s}
    \\
    \mcal P &= \imu \bar \Psi \gm^5 \Psi
    , \label{eq:db_p}
    \\
    \mcal V^\mu &= \bar \Psi \gm^\mu \Psi
    = (\rho/e, \bm j/ec)
    , \label{eq:db_v}
    \\
    \mcal A^\mu &= \bar \Psi \gm ^\mu \gm^5 \Psi
    = (-\tau^Z, \bm {\mathcal A} )
    , \label{eq:db_a}
    \\
    \mcal T^{\mu\nu} &= \imu \bar \Psi \sg^{\mu\nu} \Psi
    = (\bm{\mathcal P}, \bm{\mathcal M})
    ,\label{eq:db_t}
\end{align}
where $\sg^{\mu\nu} = \frac{1}{2}[\gm^\mu, \gm^\nu]$ is the antisymmetric tensor and $\bar\Psi = \Psi^\dg\gm^0$ is the Dirac conjugate.
Equations~\eqref{eq:db_s}-\eqref{eq:db_t} correspond to Lorentz scalar, pseudoscalar, vector, pseudovector, and tensor, respectively.
For the tensor in Eq.~\eqref{eq:db_t}, we have introduced a short-hand notation defined as
\begin{align}
( \bm{\mathcal P}, \bm{\mathcal M} )
&= 
\begin{pmatrix}
     0 & \mathcal P^x & \mathcal P^y & \mathcal P^z \\
     -\mathcal P^x & 0 & -\mathcal M^z & \mathcal M^y \\
     -\mathcal P^y & \mathcal M^z & 0 & -\mathcal M^x \\
     -\mathcal P^z & -\mathcal M^y & \mathcal M^x & 0 \\
 \end{pmatrix}
 .
\end{align}
The right-hand sides of Eqs.~\eqref{eq:db_v}-\eqref{eq:db_t} correspond to electronic charge $\rho$, electronic current $\bm j$, electron chirality $\tau^Z$, polarization $\bm{\mathcal P}$, and magnetization $\bm{\mathcal M}$.
These physical quantities characterize the electronic state in materials.

Table~\ref{tab:bilinears} summarizes the parity under spatial inversion (SI) and time reversal (TR) for the bilinears defined in Eqs.~\eqref{eq:db_s}-\eqref{eq:db_t}, categorized into scalar (one-component) and vector (three-components) quantities.
The left table represents the scalar quantities, while the right one represents the vector quantities.

Note that the systematic investigation of electronic asymmetry has also been discussed in the context of multipoles.
Particularly, electric toroidal multipoles are being recognized as a novel electric degree of freedom, in addition to conventional electric and magnetic multipoles \cite{Dubovik86, Dubovik90, Prosandeev06, Guo12, Hayami18, Hayami19, Kusunose20, Oiwa22, Hayami22, Hirose22, Kishine22, Hoshino23, Kusunose24, Inda24, Hayami24}.
It is pointed out that the electric toroidal multipole moments cannot be characterized by charge, spin, or current density, but can be characterized by the electron chirality $\tau^Z$ and spin-derived electric polarization $\bm{\mcal P}_S$ \cite{Hoshino23}, which will be introduced in the next subsection.

\subsection{Non-relativistic limit \label{sec:nrl}}

Now, we focus on low-energy behavior, which becomes important in condensed matter physics.
We take the non-relativistic limit (NRL) and represent the physical quantities in Eqs.~\eqref{eq:db_s}-\eqref{eq:db_t} by the two-component Schr\"odinger field $\psi = (\psi_\ua, \psi_\da)^\T$ ($\ua$, $\da$ is a spin degree of freedom).
The NRL of the Dirac equation is widely used in the context of condensed matter physics to discuss the relativistic corrections of the Hamiltonian, such as the Darwin term and spin-orbit coupling \cite{Berestetskii_book, Bjorken_book}.
The derivation of these terms corresponds to the $1/m$ (or $1/c$) expansion of the Dirac equation.

The NRL of physical quantities is also derived from the $1/m$ expansion, whose detailed derivation is provided in Ref.~\cite{Hoshino24}.
Here, we list the NRL of Eqs.~\eqref{eq:db_s}-\eqref{eq:db_t} within the leading order of $1/m$:
\begin{align}
    \mathcal S &\simeq \psi^\dg \psi
    , \label{eq:db_s_nrl}
    \\
    \mathcal P &\simeq -\frac{\hbar}{2mc} \nabla\cdot (\psi^\dg \bm \sg \psi)
    , \label{eq:db_p_nrl}
    \\
    \mathcal V^\mu &= (\rho/e, \bm j/ec) 
    \simeq (\psi^\dg \psi, \frac{1}{2mc}\psi^\dg \dvec{\bm p} \psi)
    , \label{eq:db_v_nrl}
    \\
    \mathcal A^\mu 
    &= (-\tau^Z, \bm {\mathcal A} )
    \simeq (-\frac{1}{2mc}\psi^\dg \dvec{\bm p} \cdot \bm \sg \psi, -\psi^\dg \bm \sg \psi), \label{eq:db_a_nrl}
    \\
    \mathcal T^{\mu\nu} 
    &= (\bm{\mathcal P}, \bm{\mathcal M})
    \simeq (\frac{\hbar}{2mc}\nabla(\psi^\dg \psi) + \bm{\mcal P}_S, -\psi^\dg \bm \sg \psi)
    \label{eq:db_t_nrl}
\end{align}
with $\bm p = -\imu\hbar\nabla$ and $A\dvec{O}B = AOB + (O^\ast A) B$.
These representations are summarized in Tab.~\ref{tab:bilinears}.
Note that the first term of $\bm{\mathcal P}$ is independent of spin and vanishes after the spatial integration.
The second term, on the other hand, is a spin-derived electric polarization given by $\bm{\mcal P}_S = -\frac{1}{2mc} \psi^\dg \dvec{\bm p} \times \bm \sg \psi$, which can be finite after the spatial integration.

DiracBilinears.jl is a tool for computing the quantities in Eqs.~\eqref{eq:db_s_nrl}-\eqref{eq:db_t_nrl} quantitatively.
Below, we explain how to evaluate these quantities using the output of external first-principles calculation packages.

\subsection{Bloch wave function and Plane wave basis \label{sec:bloch}}

We proceed to explain the evaluation of the spatial distribution for physical quantities using first-principles calculations.
In the following, we consider the electron chirality $\tau^Z$ in Eq.~\eqref{eq:db_a_nrl} as an example.
Other physical quantities in Eqs.~\eqref{eq:db_s_nrl}-\eqref{eq:db_t_nrl} can be computed in a similar manner.

To begin with, we expand the field operator by the Bloch wave function as $\psi_s(\bm r) = \sum_{n\bm k} \psi_{n\bm k}^s(\bm r) c_{n\bm k}$.
Then, the expectation value of $\tau^Z$ is expressed in the basis of the Bloch wave function as
\begin{align}
    \la \tau^Z(\bm r) \ra 
    &= \sum_{n\bm k ss'}[\psi_{n\bm k}^{s\ast}(\bm r) \dvec{\bm p} \cdot \bm \sg_{ss'} \psi_{n\bm k}^{s'}(\bm r)] f_{n\bm k}, \label{eq:tau_bloch}
\end{align}
where $f_{n\bm k}$ is an occupation function.
To connect with \texttt{Quantum ESPRESSO}, we expand the Bloch wave function using the plane wave basis:
\begin{align}
    \psi_{n\bm k}^s(\bm r) = \frac{1}{\sqrt V}\sum_{\bm G} C_{n\bm k}^s(\bm G) \epn^{\imu(\bm k + \bm G) \cdot \bm r}, \label{eq:planewave}
\end{align}
where $\bm G$ is a reciprocal lattice vector and $V$ is a system volume.
Since the $\bm r$-dependence in Eq.~\eqref{eq:planewave} is included only in the exponential factor, we can perform differentiation and integral of $\bm r$ analytically.
In the plane wave basis, Eq.~\eqref{eq:tau_bloch} is expressed as
\begin{align}
    \la \tau^Z(\bm r) \ra 
    =\frac{2}{V} \sum_{\bm k n s s' \bm G \bm G'} \real[C_{\bm k n}^{s\ast}(\bm G) \hbar (\bm k + \bm G') \cdot \bm \sg_{s s'} \nt 
    \times C_{\bm k n}^{s'}(\bm G') \epn^{- \imu (\bm G - \bm G') \cdot \bm r}] f_{n\bm k}, \label{eq:tauz_plane}
\end{align}
where the summation of $\bm k$ is taken over the Brillouin zone.

\begin{figure*}[tb]
    \centering
    \includegraphics[width=17cm]{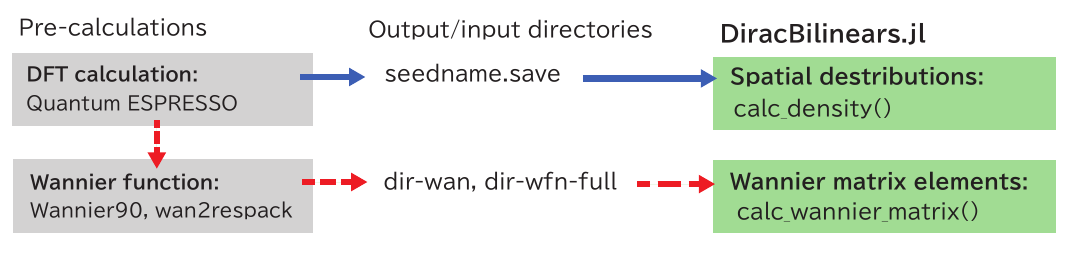}
    \caption{
    Calculation flow of Diracbilinears.jl.
    The blue solid arrows represent the calculation procedure for spatial distributions (Sec.~\ref{sec:usage_spatial}), while the red dashed arrows represent the procedure for computing Wannier matrix elements (Sec.~\ref{sec:usage_wannier}).
    These two calculations can be performed independently.
    }
    \label{fig:flow}
\end{figure*}

\subsection{Wannier interpolation \label{sec:wannier}}

Wannier interpolation takes advantage of the real-space localization of Wannier functions to achieve dense $\bm k$-mesh sampling for electronic band structures and related quantities.
In addition to the spatial distribution discussed in the previous subsection, DiracBilinears.jl can also compute the matrix elements of the Dirac bilinears in Eqs.~\eqref{eq:db_s_nrl}-\eqref{eq:db_t_nrl} in the Wannier function basis (Wannier matrix elements).

The Wannier function characterized by the lattice vector $\bm R$ is defined by
\begin{align}
    w_{m\bm R}^s(\bm r)
    = \frac{1}{\sqrt{N_{\bm k}}} \sum_{n\bm k} \psi_{n\bm k}^s(\bm r) U_{nm}^{(\bm k)} \epn^{-\imu \bm k \cdot \bm R}, \label{eq:wannier}
\end{align}
where $N_{\bm k}$ is a number of $\bm k$-point.
We introduce the plane wave expansion coefficient for the Wannier function, following the output format of \texttt{wan2respack}.
Inserting Eq.~\eqref{eq:planewave} into Eq.~\eqref{eq:wannier}, we have
\begin{align}
    w_{m\bm R}^s(\bm r) = \frac{1}{N_{\bm k}}\frac{1}{\sqrt{V_{\mrm c}}} \sum_{\bm k \bm G} \tilde C_{m \bm k}^s(\bm G) \epn^{\imu(\bm k + \bm G) \cdot \bm r} \epn^{-\imu \bm k \cdot \bm R} \label{eq:wan_c}
\end{align}
with $\tilde C_{m \bm k}^s(\bm G) = \sum_n C_{n \bm k}^s(\bm G) U_{nm}^{(\bm k)}$ and $V_{\mrm c} = V/N_{\bm k}$.
The Wannier matrix elements are often used for evaluating tight-binding parameters $H_{mm'}(\bm R) = \sum_{ss'} \int \diff\bm r w_{m\bm 0}^{s\ast}(\bm r) H_{ss'}(\bm r) w_{m'\bm R}^{s'}(\bm r)$ with Kohn-Sham Hamiltonian $H_{ss'}(\bm r)$.
Diracbilinears.jl provides the Wannier matrix elements of the Dirac bilinears in Eqs.~\eqref{eq:db_s_nrl}-\eqref{eq:db_t_nrl}.
Here, we consider the case of electron chirality $[\tau^Z(\bm R)]_{mm'} = \sum_{ss'} \int\diff\bm r w_{m\bm 0}^{s\ast}(\bm r) \dvec{\bm p} \cdot \bm \sg_{ss'} w_{m'\bm R}^{s'}(\bm r)$ as an example.
Using Eq.~\eqref{eq:wan_c}, $[\tau^Z(\bm R)]_{mm'}$ is represented as
\begin{align}
    [\tau^Z(\bm R)]_{mm'}
    &= \frac{2}{N_{\bm k}} \sum_{\bm k s s' \bm G} \tilde C^s_{\bm k m}(\bm G)^\ast \hbar(\bm k + \bm G) \cdot \bm \sg_{ss'} \tilde C^{s'}_{\bm k m'}(\bm G) \epn^{-\imu\bm k \cdot \bm R}. \label{eq:tauz_wan}
\end{align}

One application of the Wannier matrix elements in Eq.~\eqref{eq:tauz_wan} is to evaluate the total chirality, which is given by the integral of $\la \tau^Z(\bm r) \ra$ over the unit cell:
\begin{align}
    &C = \int_{\mrm{cell}}\diff\bm r\, \la \tau^Z(\bm r) \ra = \frac{1}{N_{\bm k}} \sum_{n\bm k} \tau^Z_{n\bm k} f_{n\bm k} \label{eq:total_tauz}
\end{align}
with $\tau^Z_{n\bm k} = \sum_{ss'}\int \diff\bm r \psi_{n\bm k}^{s\ast}(\bm r) \bm p \cdot \bm \sg_{ss'} \psi_{n\bm k}^{s'}(\bm r)$.
The chirality in the representation of Bloch wave function $\tau^Z_{n\bm k}$ in Eq.~\eqref{eq:total_tauz} is related to the Wannier matrix elements as follows:
\begin{align}
    \tau^Z_{n\bm k} = \sum_{mm' \bm R} \mcal{U}^{(\bm k)\ast}_{mn} [\tau^Z(\bm R)]_{mm'} \mcal{U}^{(\bm k)}_{m'n} \epn^{\imu\bm k \cdot \bm R} \label{eq:tauz_int}
\end{align}
where $\mcal{U}^{(\bm k)}_{mn}$ is a unitary matrix which diagonalizes the tight-binding Hamiltonian.
To evaluate the total chirality $C$ using Wannier interpolation, we compute $\tau^Z_{n\bm k}$ from Eq.~\eqref{eq:tauz_int} and replace $N_{\bm k}$ in Eq.~\eqref{eq:total_tauz} with the number of interpolated $\bm k$-points.

\section{Installation and usage \label{sec:usage}}

In this section, we provide the installation and usage of DiracBilinear.jl.
The source code of DiracBilinears.jl is available at Julia’s package manager by executing the following command:
\begin{lstlisting}[language=julia, label=install]
using Pkg
Pkg.add("DiracBilinears")
\end{lstlisting}

Figure~\ref{fig:flow} shows the calculation flow of DiracBilinears.jl.
We begin by performing density functional theory (DFT) calculations using \texttt{Quantum ESPRESSO}. 
To determine the spatial distribution of the physical quantities, we then use DiracBilinears.jl (indicated by the blue solid arrows in Fig.~\ref{fig:flow}).
On the other hand, to compute the Wannier matrix elements, we perform \texttt{Wannier90} and \texttt{wan2respack} and generate Wannier functions before we use DiracBilinears.jl (indicated by the red dashed arrows in Fig.~\ref{fig:flow}).
These two calculations can be performed independently.
In the following subsection, we explain the detailed usage of DiracBilinears.jl for computing spatial distribution (Sec.~\ref{sec:usage_spatial}) and Wannier matrix elements (Sec.~\ref{sec:usage_wannier}) of the Dirac bilinears.

\subsection{Spatial distribution \label{sec:usage_spatial}}

Now, we explain the calculation procedure of the spatial distribution, which is represented by the blue solid arrows in Fig.~\ref{fig:flow}.
To obtain the spatial distribution, we need the following files for \texttt{Quantum ESPRESSO} calculations:
\begin{description}
    \setlength{\parskip}{0cm}
    \setlength{\itemsep}{0.2cm}
    \item[\labelitemi\ \texttt{seedname.scf.in}] \mbox{} \\
        The input file for scf calculation.
    \item[\labelitemi\ \texttt{seedname.nscf.in}] \mbox{} \\
        The input file for nscf calculation.
\end{description}
Since the calculation of the spatial distribution requires Bloch wave functions for the full Brillouin zone, the nscf calculation needs to specify the uniform $\bm k$-point grid.

To begin with, we need to perform DFT calculations by \texttt{Quantum ESPRESSO} to obtain the Bloch wave functions.
This involves carrying out scf and nscf calculations:
\begin{lstlisting}[label=qesp]
$\mbox{\textdollar}$ QE/bin/pw.x < seedname.scf.in
                > seedname.scf.out
$\mbox{\textdollar}$ QE/bin/pw.x < seedname.nscf.in
                > seedname.nscf.out
\end{lstlisting}
This package calculates the spatial distribution using the output files of Bloch wave functions from the nscf calculation stored in the \texttt{seedname.save} directory.

We describe how to obtain the spatial distribution of the physical quantities in Eqs.~\eqref{eq:db_s_nrl}-\eqref{eq:db_t_nrl}.
To compute the spatial distribution of the electron chirality $\tau^Z(\bm r)$ in Eq.~\eqref{eq:tauz_plane}, we use the following commands:
\begin{lstlisting}[language=julia, label=space]
using DiracBilinears

$\tau$z = calc_density(calc="$\color{Maroon}{\tau}$z", 
    qedir="$\color{Maroon}{\mbox{\textdollar}}$outdir/seedname.save" [, nrmesh])
\end{lstlisting}
The physical quantities to be calculated can be selected by specifying the option \texttt{calc="$\mrm{\rho}$"} (electron density), \texttt{calc="ms"} (magnetization), \texttt{calc="j"} (current), \texttt{calc="$\mrm{\nabla ms}$"} (pseudoscalar),
\texttt{calc="$\mrm{\tau z}$"} (electron chirality), \texttt{calc="$\mrm{\nabla\rho}$"} (gradient part of polarization), or \texttt{calc="ps"} (spin-induced electric polarization).
We can also set the size of the spatial mesh by specifying the option \texttt{nrmesh}, for example, \texttt{nrmesh=(30, 30, 30)}.
If not specified, the default value is automatically determined by the FFT grid of the \texttt{nscf} calculation.
When specifying \texttt{nrmesh}, it should be chosen with reference to the FFT mesh, since the spatial distribution is computed using FFT.

The return values of \texttt{calc\_density()} are arrays whose sizes are \texttt{nrmesh} for the scalar quantities (\texttt{calc="$\mrm{\rho}$", "$\mrm{\nabla ms}$", "$\mrm{\tau z}$"}) and (3, \texttt{nrmesh}) for the vector quantities (\texttt{calc="ms", "j", "$\mrm{\nabla \rho}$", "ps"}).
For \texttt{qedir}, we need to specify the output directory of \texttt{Quantum ESPRESSO} calculation, which contains the Bloch wave function, and \texttt{\$outdir} is the directory specified in the input for the nscf calculation.
Note that the return values for current $\bm j$, pseudoscalar $-\frac{\hbar}{2mc} \nabla \cdot (\psi^\dg\bm \sg\psi)$, electron chirality $\tau^Z$, gradient part of electric polarization $\frac{\hbar}{2mc}\nabla (\psi^\dg \psi)$, and spin-derived electric polarization $\bm{\mcal P}_S$ are provided in the atomic unit, with the factor $\hbar/2mc$ omitted.

We also provide the function for \texttt{XCrySDen} plot \cite{XCrySDen}.
After we perform \texttt{calc\_density}, we can generate the xsf file by the following command:
\begin{lstlisting}[language=julia, label=space_save]
write_density($\tau$z; 
    qedir="$\color{Maroon}{\mbox{\textdollar}}$outdir/seedname.save",
    savefile=FILENAME)
\end{lstlisting}
The file extension of FILENAME must be set to ``.xsf''.

\subsection{Wannier matrix elements \label{sec:usage_wannier}}

We here explain the procedure for computing the matrix elements in the basis of the Wannier function.
For calculation, we need the following files:
\begin{description}
    \setlength{\parskip}{0cm}
    \setlength{\itemsep}{0.2cm}
    \item[\labelitemi\ \texttt{seedname.scf.in}] \mbox{} \\
        The input file for scf calculation.
    \item[\labelitemi\ \texttt{seedname.pw2wan.in}] \mbox{} \\
        The input file for \texttt{pw2wannier90.x}.
    \item[\labelitemi\ \texttt{seedname.win.ref}] \mbox{} \\
        The reference file for generating the input file of \texttt{Wannier90}.
    \item[\labelitemi\ \texttt{conf.toml}] \mbox{} \\
        The configuration file for \texttt{wan2respack}.
\end{description}
These input files were prepared following the usage of wan2respack \cite{Kurita23}.

First, we perform scf calculation by \texttt{Quantum ESPRESSO}:
\begin{lstlisting}[label=qewan]
$\mbox{\textdollar}$ QE/bin/pw.x < seedname.scf.in
                > seedname.scf.out
\end{lstlisting}
Next, we compute the Wannier functions using \texttt{Wannier90} and \texttt{wan2respack} (for more information, see Refs.~\cite{Mostofi08, Pizzi20, Kurita23}).
For the calculations using \texttt{wan2respack}, we need to use the code from the ``spinor'' branch on GitHub.
To begin with, we execute the following command:
\begin{lstlisting}[label=wan2respack1]
$\mbox{\textdollar}$ python wan2respack/bin/wan2respack.py 
   -pp conf.toml
\end{lstlisting}
After this calculation, the input file for nscf calculation (\texttt{seedname.nscf\_wannier.in}) is generated.
Note that we need to replace \texttt{calc="scf"} with \texttt{calc="nscf"} in the file.
Below, we list the remaining commands for \texttt{Wannier90} and \texttt{wan2respack}.
\begin{lstlisting}[label=wan2respack2]
$\mbox{\textdollar}$ QE/bin/pw.x < seedname.nscf_wannier.in 
                > seedname.nscf_wannier.out
$\mbox{\textdollar}$ Wanier90/wannier90.x -pp seedname
$\mbox{\textdollar}$ QE/bin/pw2wannier90.x < seedname.pw2wan.in 
                           > seedname.pw2wan.out
$\mbox{\textdollar}$ Wanier90/wannier90.x seedname
$\mbox{\textdollar}$ python wan2respack/bin/wan2respack.py 
   conf.toml
\end{lstlisting}
After the above calculations, the output directories ``dir-wfn-full'' and ``dir-wan'' are generated in the execution directory.

Now, we compute the matrix elements specifying the path of two directories dir-wfn-full (\texttt{WFNDIR}) and dir-wan (\texttt{WANDIR}), which are generated by \texttt{wan2respack} calculation.
The following command is used to calculate the matrix elements of electron chirality $[\tau^Z(\bm R)]_{mm'}$ in Eq.~\eqref{eq:tauz_wan}:
\begin{lstlisting}[language=julia, label=wan]
using DiracBilinears

rs, degen = calc_rgrid(rfile=FILENAME 
    [, mpmesh])
$\tau$z = calc_wannier_matrix(calc="$\color{Maroon}{\tau}$z", rgrid=rs, wfndir=WFNDIR, wandir=WANDIR)
\end{lstlisting}
where the return value of \texttt{calc\_wannier\_matrix()} is the matrix elements $[\tau^Z(\bm R)]_{mm'}$.
For \texttt{rfile}, we specify either the input file for \texttt{Wannier90} (\texttt{seedname.win}) or the input file for the scf calculation (\texttt{seedname.scf.in}), which is used to generate $\bm R$-grid.
The size of the $\bm{R}$-grid can also be specified using the \texttt{mpmesh} option.
We can select the type of \texttt{calc} from the following choices: \texttt{calc="$\mrm{\rho}$"} (electron density), \texttt{calc="ms"} (magnetization), \texttt{calc="j"} (current),
\texttt{calc="$\mrm{\tau z}$"} (electron chirality), and \texttt{calc="ps"} (spin-induced electric polarization).
The return values for current $\bm j$, electron chirality $\tau^Z$, and spin-derived electric polarization $\bm{\mcal P}_S$ are given in the atomic unit, with the factors $\hbar/2mc$ omitted.

The Wannier matrix elements can be saved to a \texttt{.dat} file using the following command:
\begin{lstlisting}[language=julia, label=wan_w90]
write_wannier_matrix($\tau$z, rs, degen; 
    savefile=FILENAME)
\end{lstlisting}
The format is the same as that of the hopping parameters generated by \texttt{Wannier90}.

To calculate the total electron chirality $C$ in Eq.~\eqref{eq:total_tauz}, the unitary matrix $\mcal U_{mn}^{(\bm k)}$ is required.
This matrix is obtained through the following procedure: performing a Fourier transformation with respect to $\bm R$ on the tight-binding parameters from the \texttt{seedname\_hr.dat} file (an output of the \texttt{Wannier90} calculation), and then diagonalizing the resulting matrix.
We can calculate $\tau^Z_{n\bm k}$ by using Eq.~\eqref{eq:tauz_int}, and obtain $C$ by summing over $\bm k$ and $n$ as described in Eq.~\eqref{eq:total_tauz}.

\section{Application \label{sec:application}}

\begin{figure}[tb]
    \centering
    \includegraphics[width=8.5cm]{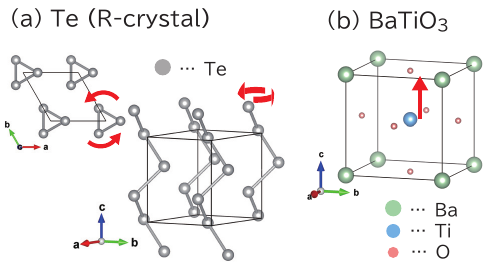}
    \caption{
    Crystal structure of right-handed Te and $\mrm{BaTiO_3}$.
    The crystal Te has the helical structure, and $\mrm{BaTiO_3}$ has a Ti atom displaced from the center of the surrounding O atoms, as indicated by the red arrows.
    }
    \label{fig:crystal}
\end{figure}

In this section, we explain applications of DiracBilinears.jl, presenting computational results for typical materials.
As examples, we show the results for the two cases, chirality $\tau^Z(\bm r)$ for the chiral crystal Te and polarization $\bm{\mcal P}(\bm r)$ for polar crystal ${\rm BaTiO_3}$.

Figure~\ref{fig:crystal} shows the crystal structure of (a) right-handed Te and (b) $\mrm{BaTiO_3}$.
The crystal structure of Te has three-fold rotational axes along the $c$-axis, located at the corners of the unit cell, which are outlined by black lines in the figure. 
The atoms are helically arranged around these rotational axes, as indicated by the red arrows in Fig.~\ref{fig:crystal} (a).
In the crystal $\mrm{BaTiO_3}$, the Ti atom is displaced along the $c$-axis from the center of the surrounding O atoms, forming a polar structure as indicated by the red arrow in Fig.~\ref{fig:crystal} (b).

In \texttt{Quantum ESPRESSO} calculations, we use Perdew–Burke–Ernzerhof (PBE) exchange-correlation functional \cite{Perdew96}, and optimized norm-conserving Vanderbilt pseudopotential \cite{Hamann13} provided in \texttt{PseudoDojo} \cite{vanSetten18}.
The calculations are carried out with $6 \times 6 \times 6$ $\bm{k}$-mesh, and the plane wave cutoff energies set to $70\, \mathrm{Ry}$ for Te and $90\, \mathrm{Ry}$ for ${\rm BaTiO_3}$ (for both scf and nscf calculations).
We use the Methfessel-Paxton first-order spreading \cite{Methfessel89} with the width of $0.01\, \mrm{Ry}$ for smearing in \texttt{Quantum ESPRESSO} calculation, the evaluation of Eq.~\eqref{eq:total_tauz}, and similar calculation for the electronic polarization.
The sample input files are available in the GitHub repository \footnote{\url{https://github.com/TatsuyaMiki/DiracBilinears.jl.git}}.

\subsection{Te (chiral crystal)}

\begin{figure}[tb]
    \centering
    \includegraphics[width=8.0cm]{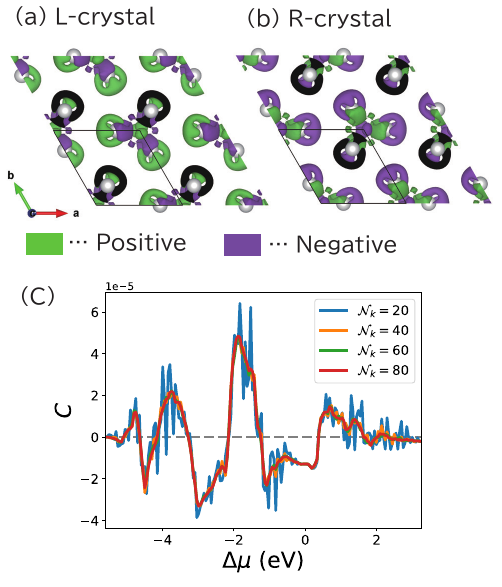}
    \caption{Spatial distribution of electron chirality for (a) left-handed Te and (b) right-handed Te. 
    (c) Chemical potential dependence of total electron chirality for R-crystal Te.
    The lines correspond to calculations with different interpolated $\bm k$-mesh sizes.
    We use the same lattice constant as that employed in Ref.~\cite{Miki24}.
    The black shades in (a) and (b) are the cross-section of the cell.}
    \label{fig:te}
\end{figure}

Te crystal has a simple chiral structure consisting of three Te atoms and is a suitable material to demonstrate the calculation of $\tau^Z(\bm r)$.
It exists in two enantiomeric configurations, the left-handed (L-) and right-handed (R-) crystals, which are related to each other by inversion symmetry.

Figures~\ref{fig:te} (a) and (b) show the spatial distribution of electron chirality, which can be calculated following the procedure explained in Sec.~\ref{sec:usage_spatial}.
The distribution for the L-crystal is presented in Fig.~\ref{fig:te} (a), and that for the R-crystal is shown in Fig.~\ref{fig:te} (b).
The green (purple) area indicates the positive (negative) values.
Since the atomic positions of the L-crystal and R-crystal are connected by spatial inversion, the electron chirality (pseudo-scalar) has opposite signs in the two crystals.

We turn our attention to the results of Wannier matrix elements explained in Sec.~\ref{sec:usage_wannier}.
We generate the Wannier functions for the $p$-orbital subspace located around the Fermi energy by using \texttt{Wannier90} and \texttt{wan2respack}.
After performing the calculations described in Sec.~\ref{sec:usage_wannier}, we can obtain the matrix element $[\tau^Z(\bm R)]_{mm'}$ from the output of \texttt{write\_wannier\_matrix()}.
Table~\ref{tab:wannier} shows the list of the matrix element $[\tau^Z(\bm R)]_{mm'}$ sorted by the absolute values, where the values in the table are displayed to five decimal places.

The total electron chirality $C$ given in Eq.~\eqref{eq:total_tauz} is evaluated by using $[\tau^Z(\bm R)]_{mm'}$ and \texttt{seedname\_hr.dat} generated by the \texttt{Wannier90} calculation.
It can be computed by applying Eq.\eqref{eq:tauz_int} as discussed in Sec.~\ref{sec:usage_wannier}.
Figure~\ref{fig:te} (c) shows the chemical potential dependence of $C$ for several interpolated $\bm{k}$-mesh sizes.
Each line in Fig~\ref{fig:te} (c) represents the results calculated on $\mcal N_{\bm k} \times \mcal N_{\bm k} \times \mcal N_{\bm k}$ interpolated $\bm k$-mesh.
The curves become smoother and the values converge as $\mcal N_{\bm k}$ increases.

\begin{table}[tb]
    \centering
    \begin{tabular}{c|c|r}
        $\bm R = [n_1, n_2, n_3]$ & ($m, m'$) & $[\tau^Z(\bm R)]_{mm'}$ \\ \hline
        $[1, 1, -1]$ & $(12, 13)$ & $0.39607 + 0.00846 \imu$ \\
        $[-1, -1, 1]$ & $(13, 12)$ & $0.39607 -0.00846\imu$ \\
        $[1, 1, -1]$ & $(11, 14)$ & $-0.39607 + 0.00846\imu$ \\
        $[-1, -1, 1]$ & $(14, 11)$ & $-0.39607 -0.00846\imu$ \\
        $[1, 1, -1]$ & $(7, 18)$ & $-0.39607 -0.00846\imu$ \\
        $[-1, -1, 1]$ & $(18, 7)$ & $-0.39607 + 0.00846\imu$ \\
        $[1, 1, -1]$ & $(8, 17)$ & $0.39607 -0.00846\imu$ \\
        $[-1, -1, 1]$ & $(17, 8)$ & $0.39607 + 0.00846\imu$ \\
        $[1, 1, -1]$ & $(7, 17)$ & $-0.00028 + 0.37186\imu$ \\
        $[-1, -1, 1]$ & $(17, 7)$ & $-0.00028 -0.37186\imu$ \\
        \vdots & \vdots & \vdots
    \end{tabular}
    \caption{
    The Wannier matrix elements of electron chirality, $[\tau^Z(\bm R)]_{mm'}$, given in Eq.~\eqref{eq:tauz_wan} for R-crystal Te. The left column represents the lattice vector $\bm R$, the center column lists the indices $m, m'$, and the right column shows the corresponding matrix elements. 
    The values are sorted by their absolute values and displayed to five decimal places.
    }
    \label{tab:wannier}
\end{table}

\subsection{${\rm BaTiO_3}$ (polar crystal)}

\begin{figure}[tb]
    \centering
    \includegraphics[width=7cm]{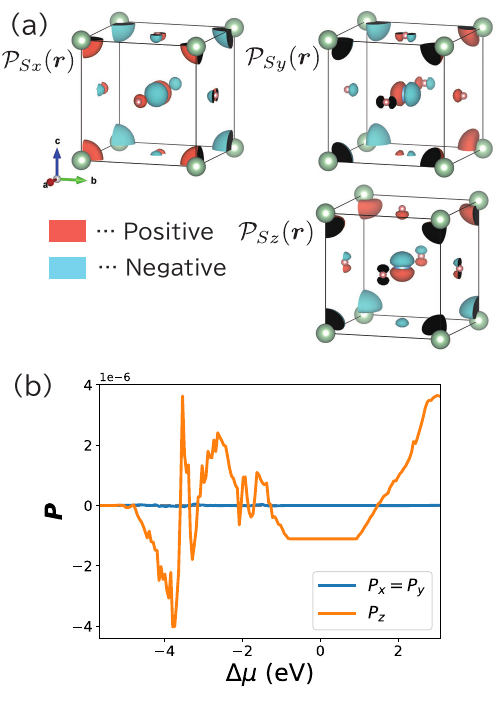}
    \caption{(a) Spatial distribution and (b) total value of spin-derived electric polarization.
    We use the same lattice constant as that employed in Ref.~\cite{Miki24}.
    The black shades in (a) are the cross-section of the cell.}
    \label{fig:batio3}
\end{figure}

${\rm BaTiO_3}$ is a typical polar crystal in which the Ti atom is displaced along the $c$-axis from the center of the surrounding O atoms.
In this subsection, we present the results for the electric polarization $\bm{\mcal P} = \frac{1}{4mc}\nabla(\psi^\dg\psi) + \bm{\mcal P}_S$ defined in Eq.~\eqref{eq:db_t_nrl}, with ${\rm BaTiO_3}$ as an example.
Since the first term in $\bm{\mcal P}$, corresponding to the gradient of the electronic density, vanishes by the spatial integration, we focus on the spin-derived electric polarization $\bm{\mcal P}_S$ in the following.

Here, we focus on the spatial distribution, which we calculate following the procedure demonstrated in Sec.~\ref{sec:usage_spatial}, with the option \texttt{calc=ps}.
Figure~\ref{fig:batio3}(a) shows the spatial distribution of $\la\bm{\mcal P}_S(\bm r)\ra$, where red (blue) regions indicate positive (negative) values. 
Each component $\la\mcal P_{Si}(\bm r)\ra\ (i=x, y, z)$ is polarized along the $i$-axis, with positive and negative values distributed in the $i$-direction.

Additionally, we present the results of the Wannier matrix elements. 
The Wannier orbitals consist of $p$-orbitals for the O atoms and $d_{xy}$-, $d_{yz}$-, $d_{xz}$-orbitals for the Ti atom. 
The total electric polarization $\bm P = \int_{\mrm{cell}} \diff\bm r \la \bm{\mcal P}_S(\bm r) \ra$ can be computed in a manner similar to the case of electron chirality, which is discussed in Sec.~\ref{sec:usage_wannier}.
The calculation can be performed by specifying the option \texttt{calc=ps}.
Figure~\ref{fig:batio3}(b) shows the chemical potential dependence of the total polarization $\bm P$. 
For this plot, we set the interpolated $\bm k$-mesh size to $\mcal N_{\bm k} = 100$. Since the crystal structure of $\mrm{BaTiO_3}$ has a rotational symmetry along the $z$-axis, the $x$- and $y$-components $P_x$ and $P_y$ vanish (blue line) and only $P_z$ (orange line) remains non-zero.

\section{Summary \label{sec:summary}}

In this paper, we have introduced DiracBilinears.jl, a Julia package developed for computing Dirac bilinears derived from relativistic quantum theory.
To focus on the low energy, we have taken the non-relativistic limit of the bilinears.
This package offers tools to compute the spatial distribution and the Wannier matrix elements of the bilinears.
It interfaces with first-principles calculation packages such as \texttt{Quantum ESPRESSO}, \texttt{Wannier90}, and \texttt{wan2respack}, providing a practical framework for exploring relativistic effects in materials.

As a demonstration of DiracBilinears.jl, we have evaluated the spatial distribution of electron chirality for the chiral crystal Te and the spin-induced polarization for the polar crystal ${\rm BaTiO_3}$.
Additionally, we provide the chemical potential dependence of these total values in crystals as an application of the Wannier matrix elements.

This paper has focused on the evaluation of Dirac bilinears, taking into account the leading order of $1/m$.
On the other hand, future advancements of this package may include evaluating higher-order relativistic corrections to the bilinears and other physical quantities discussed in Ref.~\cite{Hoshino24}, such as chirality polarization appearing in the time evolution equation of the axial current.

\section*{Acknowledgments}

TM is grateful to H. Ikeda, M.-T. Suzuki, and S. Hoshino for the valuable discussions.
He also appreciates M. Fukuda for his beneficial comments, which helped improve this paper.
This work was supported by the KAKENHI Grants No.~23KJ0298 (TM), No.~23H04869 (TK, YN), and No.~22K03447 (TK).



%% References


\bibliographystyle{elsarticle-num}
\bibliography{cmptchiral}

\begin{thebibliography}{10}
\expandafter\ifx\csname url\endcsname\relax
  \def\url#1{\texttt{#1}}\fi
\expandafter\ifx\csname urlprefix\endcsname\relax\def\urlprefix{URL }\fi
\expandafter\ifx\csname href\endcsname\relax
  \def\href#1#2{#2} \def\path#1{#1}\fi

\bibitem{Naaman15}
R.~Naaman, D.~H. Waldeck, Spintronics and chirality: Spin selectivity in electron transport through chiral molecules, Annu. Rev. Phys. Chem. 66 (2015) 263--281.
\newblock \href {https://doi.org/10.1146/annurev-physchem-040214-121554} {\path{doi:10.1146/annurev-physchem-040214-121554}}.

\bibitem{Tokura18}
Y.~Tokura, N.~Nagaosa, Nonreciprocal responses from non-centrosymmetric quantum materials, Nat. Commun. 9 (2018) 3740.
\newblock \href {https://doi.org/10.1038/s41467-018-05759-4} {\path{doi:10.1038/s41467-018-05759-4}}.

\bibitem{Naaman20}
R.~Naaman, Y.~Paltiel, D.~H. Waldeck, Chiral induced spin selectivity gives a new twist on spin-control in chemistry, Acc. Chem. Res. 53~(11) (2020) 2659--2667.
\newblock \href {https://doi.org/10.1021/acs.accounts.0c00485} {\path{doi:10.1021/acs.accounts.0c00485}}.

\bibitem{Tokura21}
Y.~Tokura, N.~Kanazawa, Magnetic skyrmion materials, Chem. Rev. 121~(5) (2021) 2857--2897.
\newblock \href {https://doi.org/10.1021/acs.chemrev.0c00297} {\path{doi:10.1021/acs.chemrev.0c00297}}.

\bibitem{Bloom24}
B.~P. Bloom, Y.~Paltiel, R.~Naaman, D.~H. Waldeck, Chiral induced spin selectivity, Chem. Rev. 124~(4) (2024) 1950--1991.
\newblock \href {https://doi.org/10.1021/acs.chemrev.3c00661} {\path{doi:10.1021/acs.chemrev.3c00661}}.

\bibitem{Gohler11}
B.~G\"ohler, V.~Hamelbeck, T.~Z. Markus, M.~Kettner, G.~F. Hanne, Z.~Vager, R.~Naaman, H.~Zacharias, Spin selectivity in electron transmission through self-assembled monolayers of double-stranded dna, Science 331~(6019) (2011) 894--897.
\newblock \href {https://doi.org/10.1126/science.1199339} {\path{doi:10.1126/science.1199339}}.

\bibitem{Brinkman24}
S.~S. Brinkman, X.~L. Tan, B.~Brekke, A.~C. Mathisen, O.~Finnseth, R.~J. Schenk, K.~Hagiwara, M.-J. Huang, J.~Buck, M.~Kall\"ane, M.~Hoesch, K.~Rossnagel, K.-H. Ou~Yang, M.-T. Lin, G.-J. Shu, Y.-J. Chen, C.~Tusche, H.~Bentmann, Chirality-driven orbital angular momentum and circular dichroism in cosi, Phys. Rev. Lett. 132 (2024) 196402.
\newblock \href {https://doi.org/10.1103/PhysRevLett.132.196402} {\path{doi:10.1103/PhysRevLett.132.196402}}.

\bibitem{Sakano20}
M.~Sakano, M.~Hirayama, T.~Takahashi, S.~Akebi, M.~Nakayama, K.~Kuroda, K.~Taguchi, T.~Yoshikawa, K.~Miyamoto, T.~Okuda, K.~Ono, H.~Kumigashira, T.~Ideue, Y.~Iwasa, N.~Mitsuishi, K.~Ishizaka, S.~Shin, T.~Miyake, S.~Murakami, T.~Sasagawa, T.~Kondo, Radial spin texture in elemental tellurium with chiral crystal structure, Phys. Rev. Lett. 124 (2020) 136404.
\newblock \href {https://doi.org/10.1103/PhysRevLett.124.136404} {\path{doi:10.1103/PhysRevLett.124.136404}}.

\bibitem{Inui20}
A.~Inui, R.~Aoki, Y.~Nishiue, K.~Shiota, Y.~Kousaka, H.~Shishido, D.~Hirobe, M.~Suda, J.-i. Ohe, J.-i. Kishine, H.~M. Yamamoto, Y.~Togawa, Chirality-induced spin-polarized state of a chiral crystal ${\mathrm{crnb}}_{3}{\mathrm{s}}_{6}$, Phys. Rev. Lett. 124 (2020) 166602.
\newblock \href {https://doi.org/10.1103/PhysRevLett.124.166602} {\path{doi:10.1103/PhysRevLett.124.166602}}.

\bibitem{Rao19}
Z.~Rao, H.~Li, T.~Zhang, S.~Tian, C.~Li, B.~Fu, C.~Tang, L.~Wang, Z.~Li, W.~Fan, J.~Li, Y.~Huang, Z.~Liu, Y.~Long, C.~Fang, H.~Weng, Y.~Shi, H.~Lei, Y.~Sun, T.~Qian, H.~Ding, Observation of unconventional chiral fermions with long fermi arcs in cosi, Nature 567~(7749) (2019) 496--499.
\newblock \href {https://doi.org/10.1038/s41586-019-1031-8} {\path{doi:10.1038/s41586-019-1031-8}}.

\bibitem{Yamagishi23}
S.~Yamagishi, T.~Hayashida, R.~Misawa, K.~Kimura, M.~Hagihala, T.~Murata, S.~Hirose, T.~Kimura, Ferroaxial transitions in glaserite-type compounds: Database screening, phonon calculations, and experimental verification, Chem. Mater. 35~(2) (2023) 747--754.
\newblock \href {https://doi.org/10.1021/acs.chemmater.2c03540} {\path{doi:10.1021/acs.chemmater.2c03540}}.

\bibitem{Zhang15}
X.~Zhang, B.~Lu, R.~Li, C.~Fan, Z.~Liang, P.~Han, Structural, electronic and optical properties of ilmenite atio3(a=fe, co, ni), Mater. Sci. Semicond. Process. 39 (2015) 6--16.
\newblock \href {https://doi.org/https://doi.org/10.1016/j.mssp.2015.04.041} {\path{doi:https://doi.org/10.1016/j.mssp.2015.04.041}}.

\bibitem{Hayashida20}
T.~Hayashida, Y.~Uemura, K.~Kimura, S.~Matsuoka, D.~Morikawa, S.~Hirose, K.~Tsuda, T.~Hasegawa, T.~Kimura, Visualization of ferroaxial domains in an order-disorder type ferroaxial crystal, Nat. Commun. 11 (2020) 4582.
\newblock \href {https://doi.org/10.1038/s41467-020-18408-6} {\path{doi:10.1038/s41467-020-18408-6}}.

\bibitem{Hayashida21}
T.~Hayashida, Y.~Uemura, K.~Kimura, S.~Matsuoka, M.~Hagihala, S.~Hirose, H.~Morioka, T.~Hasegawa, T.~Kimura, Phase transition and domain formation in ferroaxial crystals, Phys. Rev. Mater. 5 (2021) 124409.
\newblock \href {https://doi.org/10.1103/PhysRevMaterials.5.124409} {\path{doi:10.1103/PhysRevMaterials.5.124409}}.

\bibitem{Hayashida23}
T.~Hayashida, K.~Kimura, T.~Kimura, Electric field–induced magnetochiral dichroism in a ferroaxial crystal, Proc. Nat. Acad. Sci. USA 120~(34) (2023) e2303251120.
\newblock \href {https://doi.org/10.1073/pnas.2303251120} {\path{doi:10.1073/pnas.2303251120}}.

\bibitem{Jin20}
W.~Jin, E.~Drueke, S.~Li, A.~Admasu, R.~Owen, M.~Day, K.~Sun, S.-W. Cheong, L.~Zhao, Observation of a ferro-rotational order coupled with second-order nonlinear optical fields, Nat. Phys. 16 (2020) 42–46.
\newblock \href {https://doi.org/10.1038/s41567-019-0695-1} {\path{doi:10.1038/s41567-019-0695-1}}.

\bibitem{Sayantika24}
B.~Sayantika, A.~S. Nicola, Electric toroidal dipole order and hidden spin polarization in ferroaxial materials (2024).
\newblock \href {http://arxiv.org/abs/2407.08369} {\path{arXiv:2407.08369}}.

\bibitem{Hlinka16}
J.~Hlinka, J.~Privratska, P.~Ondrejkovic, V.~Janovec, Symmetry guide to ferroaxial transitions, Phys. Rev. Lett. 116 (2016) 177602.
\newblock \href {https://doi.org/10.1103/PhysRevLett.116.177602} {\path{doi:10.1103/PhysRevLett.116.177602}}.

\bibitem{Hoshino23}
S.~Hoshino, M.-T. Suzuki, H.~Ikeda, Spin-derived electric polarization and chirality density inherent in localized electron orbitals, Phys. Rev. Lett. 130 (2023) 256801.
\newblock \href {https://doi.org/10.1103/PhysRevLett.130.256801} {\path{doi:10.1103/PhysRevLett.130.256801}}.

\bibitem{Hoshino24}
S.~Hoshino, T.~Miki, M.-T. Suzuki, H.~Ikeda, Dirac bilinears in condensed matter physics: Relativistic correction for observables and conjugate electromagnetic fields (2024).
\newblock \href {http://arxiv.org/abs/2408.16983} {\path{arXiv:2408.16983}}.

\bibitem{Miki24}
T.~Miki, H.~Ikeda, M.-T. Suzuki, S.~Hoshino, Quantification of electronic asymmetry: chirality and axiality in solids (2024).
\newblock \href {http://arxiv.org/abs/2410.23549} {\path{arXiv:2410.23549}}.

\bibitem{Berestetskii_book}
V.~B. Berestetskii, E.~Lifshitz, L.~P. Pitaevskii, Quantum {E}lectrodynamics, Butterworth-Heinemann, Oxford, 1982.

\bibitem{Sakurai_book}
J.~J. Sakurai, Advanced {Q}uantum {M}echanics, Addison Wesley, Reading, MA, 1967.

\bibitem{Senami19}
M.~Senami, K.~Ito, Asymmetry of electron chirality between enantiomeric pair molecules and the origin of homochirality in nature, Phys. Rev. A 99 (2019) 012509.
\newblock \href {https://doi.org/10.1103/PhysRevA.99.012509} {\path{doi:10.1103/PhysRevA.99.012509}}.

\bibitem{Kuroda22}
N.~Kuroda, T.~Oho, M.~Senami, A.~Sunaga, Enhancement of the parity-violating energy difference of ${\mathrm{h}}_{2}{X}_{2}$ molecules by electronic excitation, Phys. Rev. A 105 (2022) 012820.
\newblock \href {https://doi.org/10.1103/PhysRevA.105.012820} {\path{doi:10.1103/PhysRevA.105.012820}}.

\bibitem{Tachibana12}
A.~Tachibana, General relativistic symmetry of electron spin torque, J. Math. Chem. 50~(4) (2012) 669.
\newblock \href {https://doi.org/10.1007/s10910-011-9943-z} {\path{doi:10.1007/s10910-011-9943-z}}.

\bibitem{Hara12}
T.~Hara, M.~Senami, A.~Tachibana, Electron spin torque in atoms, Phys. Lett. A 376~(17) (2012) 1434.
\newblock \href {https://doi.org/https://doi.org/10.1016/j.physleta.2012.03.028} {\path{doi:https://doi.org/10.1016/j.physleta.2012.03.028}}.

\bibitem{Fukuda16}
M.~Fukuda, K.~Soga, M.~Senami, A.~Tachibana, Local physical quantities for spin based on the relativistic quantum field theory in molecular systems, Int. J. Quantum Chem. 116~(12) (2016) 920--931.
\newblock \href {https://doi.org/https://doi.org/10.1002/qua.25102} {\path{doi:https://doi.org/10.1002/qua.25102}}.

\bibitem{QEDalpha}
M.~Senami, K.~Ichikawa, A.~Tachibana, M.~Fukuda, K.~Soga, \href{https://github.com/mfukudaQED/QEDalpha}{\texttt{QEDalpha} package}, \url{https://github.com/mfukudaQED/QEDalpha}.
\newline\urlprefix\url{https://github.com/mfukudaQED/QEDalpha}

\bibitem{Dirac13}
\texttt{DIRAC}, a relativistic ab initio electronic structure program, Release {DIRAC13} (2013), written by L.~Visscher, H.~J.~{\relax Aa}.~Jensen, R.~Bast, and T.~Saue, with contributions from V.~Bakken, K.~G.~Dyall, S.~Dubillard, U.~Ekstr{\"o}m, E.~Eliav, T.~Enevoldsen, E.~Fa{\ss}hauer, T.~Fleig, O.~Fossgaard, A.~S.~P.~Gomes, T.~Helgaker, J.~K.~L{\ae}rdahl, Y.~S.~Lee, J.~Henriksson, M.~Ilia{\v{s}}, Ch.~R.~Jacob, S.~Knecht, S.~Komorovsk{\'y}, O.~Kullie, C.~V.~Larsen, H.~S.~Nataraj, P.~Norman, G.~Olejniczak, J.~Olsen, Y.~C.~Park, J.~K.~Pedersen, M.~Pernpointner, K.~Ruud, P.~Sa{\l}ek, B.~Schimmelpfennig, J.~Sikkema, A.~J.~Thorvaldsen, J.~Thyssen, J.~van~Stralen, S.~Villaume, O.~Visser, T.~Winther, and S.~Yamamoto (see \url{http://www.diracprogram.org}).

\bibitem{Giannozzi09}
P.~Giannozzi, S.~Baroni, N.~Bonini, M.~Calandra, R.~Car, C.~Cavazzoni, D.~Ceresoli, G.~L. Chiarotti, M.~Cococcioni, I.~Dabo, A.~D. Corso, S.~de~Gironcoli, S.~Fabris, G.~Fratesi, R.~Gebauer, U.~Gerstmann, C.~Gougoussis, A.~Kokalj, M.~Lazzeri, L.~Martin-Samos, N.~Marzari, F.~Mauri, R.~Mazzarello, S.~Paolini, A.~Pasquarello, L.~Paulatto, C.~Sbraccia, S.~Scandolo, G.~Sclauzero, A.~P. Seitsonen, A.~Smogunov, P.~Umari, R.~M. Wentzcovitch, Quantum espresso: a modular and open-source software project for quantum simulations of materials, J. Phys.: Condens. Matter 21~(39) (2009) 395502.
\newblock \href {https://doi.org/10.1088/0953-8984/21/39/395502} {\path{doi:10.1088/0953-8984/21/39/395502}}.

\bibitem{Mostofi08}
A.~A. Mostofi, J.~R. Yates, Y.-S. Lee, I.~Souza, D.~Vanderbilt, N.~Marzari, wannier90: A tool for obtaining maximally-localised wannier functions, Comput. Phys. Commun. 178~(9) (2008) 685--699.
\newblock \href {https://doi.org/https://doi.org/10.1016/j.cpc.2007.11.016} {\path{doi:https://doi.org/10.1016/j.cpc.2007.11.016}}.

\bibitem{Pizzi20}
G.~Pizzi, V.~Vitale, R.~Arita, S.~Blügel, F.~Freimuth, G.~Géranton, M.~Gibertini, D.~Gresch, C.~Johnson, T.~Koretsune, J.~Ibañez-Azpiroz, H.~Lee, J.-M. Lihm, D.~Marchand, A.~Marrazzo, Y.~Mokrousov, J.~I. Mustafa, Y.~Nohara, Y.~Nomura, L.~Paulatto, S.~Poncé, T.~Ponweiser, J.~Qiao, F.~Thöle, S.~S. Tsirkin, M.~Wierzbowska, N.~Marzari, D.~Vanderbilt, I.~Souza, A.~A. Mostofi, J.~R. Yates, Wannier90 as a community code: new features and applications, J. Phys.: Condens. Matter 32~(16) (2020) 165902.
\newblock \href {https://doi.org/10.1088/1361-648X/ab51ff} {\path{doi:10.1088/1361-648X/ab51ff}}.

\bibitem{Kurita23}
K.~Kurita, T.~Misawa, K.~Yoshimi, K.~Ido, T.~Koretsune, Interface tool from wannier90 to respack: wan2respack, Comput. Phys. Commun. 292 (2023) 108854.
\newblock \href {https://doi.org/https://doi.org/10.1016/j.cpc.2023.108854} {\path{doi:https://doi.org/10.1016/j.cpc.2023.108854}}.

\bibitem{FLPQ}
M.~Fukuda, \href{https://github.com/mfukudaQED/FLPQ}{\texttt{FLPQ} package}, \url{https://github.com/mfukudaQED/FLPQ}.
\newline\urlprefix\url{https://github.com/mfukudaQED/FLPQ}

\bibitem{OpenMX}
\textsc{OpenMX}, \url{https://www.openmx-square.org}.

\bibitem{Ozaki03}
T.~Ozaki, Variationally optimized atomic orbitals for large-scale electronic structures, Phys. Rev. B 67 (2003) 155108.
\newblock \href {https://doi.org/10.1103/PhysRevB.67.155108} {\path{doi:10.1103/PhysRevB.67.155108}}.

\bibitem{Ozaki04}
T.~Ozaki, H.~Kino, Numerical atomic basis orbitals from h to kr, Phys. Rev. B 69 (2004) 195113.
\newblock \href {https://doi.org/10.1103/PhysRevB.69.195113} {\path{doi:10.1103/PhysRevB.69.195113}}.

\bibitem{Dubovik86}
V.~Dubovik, L.~Tosunyan, V.~Tugushev, Axial toroidal moments in electrodynamics and solid-state physics, Sov. Phys. JETP (1986) 344.

\bibitem{Dubovik90}
V.~Dubovik, V.~Tugushev, Toroid moments in electrodynamics and solid-state physics, Phys. Rep. 187~(4) (1990) 145--202.
\newblock \href {https://doi.org/https://doi.org/10.1016/0370-1573(90)90042-Z} {\path{doi:https://doi.org/10.1016/0370-1573(90)90042-Z}}.

\bibitem{Prosandeev06}
S.~Prosandeev, I.~Ponomareva, I.~Kornev, I.~Naumov, L.~Bellaiche, Controlling toroidal moment by means of an inhomogeneous static field: An ab initio study, Phys. Rev. Lett. 96 (2006) 237601.
\newblock \href {https://doi.org/10.1103/PhysRevLett.96.237601} {\path{doi:10.1103/PhysRevLett.96.237601}}.

\bibitem{Guo12}
L.~Y. Guo, M.~H. Li, Q.~W. Ye, B.~X. Xiao, H.~L. Yang, Electric toroidal dipole response in split-ring resonator metamaterials, Eur. Phys. J. B 85~(6) (2012) 208.
\newblock \href {https://doi.org/10.1140/epjb/e2012-20935-3} {\path{doi:10.1140/epjb/e2012-20935-3}}.

\bibitem{Hayami18}
S.~Hayami, M.~Yatsushiro, Y.~Yanagi, H.~Kusunose, Classification of atomic-scale multipoles under crystallographic point groups and application to linear response tensors, Phys. Rev. B 98 (2018) 165110.
\newblock \href {https://doi.org/10.1103/PhysRevB.98.165110} {\path{doi:10.1103/PhysRevB.98.165110}}.

\bibitem{Hayami19}
S.~Hayami, Y.~Yanagi, H.~Kusunose, Y.~Motome, Electric toroidal quadrupoles in the spin-orbit-coupled metal ${\mathrm{cd}}_{2}{\mathrm{re}}_{2}{\mathrm{o}}_{7}$, Phys. Rev. Lett. 122 (2019) 147602.
\newblock \href {https://doi.org/10.1103/PhysRevLett.122.147602} {\path{doi:10.1103/PhysRevLett.122.147602}}.

\bibitem{Kusunose20}
H.~Kusunose, R.~Oiwa, S.~Hayami, Complete multipole basis set for single-centered electron systems, Journal of the Physical Society of Japan 89~(10) (2020) 104704.
\newblock \href {https://doi.org/10.7566/JPSJ.89.104704} {\path{doi:10.7566/JPSJ.89.104704}}.

\bibitem{Oiwa22}
R.~Oiwa, H.~Kusunose, Rotation, electric-field responses, and absolute enantioselection in chiral crystals, Phys. Rev. Lett. 129 (2022) 116401.
\newblock \href {https://doi.org/10.1103/PhysRevLett.129.116401} {\path{doi:10.1103/PhysRevLett.129.116401}}.

\bibitem{Hayami22}
S.~Hayami, R.~Oiwa, H.~Kusunose, Electric ferro-axial moment as nanometric rotator and source of longitudinal spin current, J. Phys. Soc. Jpn. 91~(11) (2022) 113702.
\newblock \href {https://doi.org/10.7566/JPSJ.91.113702} {\path{doi:10.7566/JPSJ.91.113702}}.

\bibitem{Hirose22}
H.~T. Hirose, T.~Terashima, D.~Hirai, Y.~Matsubayashi, N.~Kikugawa, D.~Graf, K.~Sugii, S.~Sugiura, Z.~Hiroi, S.~Uji, Electronic states of metallic electric toroidal quadrupole order in ${\mathrm{cd}}_{2}{\mathrm{re}}_{2}{\mathrm{o}}_{7}$ determined by combining quantum oscillations and electronic structure calculations, Phys. Rev. B 105 (2022) 035116.
\newblock \href {https://doi.org/10.1103/PhysRevB.105.035116} {\path{doi:10.1103/PhysRevB.105.035116}}.

\bibitem{Kishine22}
J.-i. Kishine, H.~Kusunose, H.~M. Yamamoto, On the definition of chirality and enantioselective fields, Israel Journal of Chemistry 62~(11-12) (2022) e202200049.
\newblock \href {https://doi.org/https://doi.org/10.1002/ijch.202200049} {\path{doi:https://doi.org/10.1002/ijch.202200049}}.

\bibitem{Kusunose24}
H.~Kusunose, J.-i. Kishine, H.~M. Yamamoto, {Emergence of chirality from electron spins, physical fields, and material-field composites}, Appl. Phys. Lett. 124~(26) (2024) 260501.
\newblock \href {https://doi.org/10.1063/5.0214919} {\path{doi:10.1063/5.0214919}}.

\bibitem{Inda24}
A.~Inda, R.~Oiwa, S.~Hayami, H.~M. Yamamoto, H.~Kusunose, {Quantification of chirality based on electric toroidal monopole}, J. Chem. Phys. 160~(18) (2024) 184117.
\newblock \href {https://doi.org/10.1063/5.0204254} {\path{doi:10.1063/5.0204254}}.

\bibitem{Hayami24}
S.~Hayami, H.~Kusunose, Unified description of electronic orderings and cross correlations by complete multipole representation, J. Phys. Soc. Jpn. 93~(7) (2024) 072001.
\newblock \href {https://doi.org/10.7566/JPSJ.93.072001} {\path{doi:10.7566/JPSJ.93.072001}}.

\bibitem{Bjorken_book}
J.~D. Bjorken, S.~D. Drell, Relativistic {Q}uantum {M}echanics, McGraw-Hill, New York, 1998.

\bibitem{XCrySDen}
A.~Kokalj, Xcrysden—a new program for displaying crystalline structures and electron densities, Journal of Molecular Graphics and Modelling 17~(3) (1999) 176--179.
\newblock \href {https://doi.org/https://doi.org/10.1016/S1093-3263(99)00028-5} {\path{doi:https://doi.org/10.1016/S1093-3263(99)00028-5}}.

\bibitem{Perdew96}
J.~P. Perdew, K.~Burke, M.~Ernzerhof, Generalized gradient approximation made simple, Phys. Rev. Lett. 77 (1996) 3865--3868.
\newblock \href {https://doi.org/10.1103/PhysRevLett.77.3865} {\path{doi:10.1103/PhysRevLett.77.3865}}.

\bibitem{Hamann13}
D.~R. Hamann, Optimized norm-conserving vanderbilt pseudopotentials, Phys. Rev. B 88 (2013) 085117.
\newblock \href {https://doi.org/10.1103/PhysRevB.88.085117} {\path{doi:10.1103/PhysRevB.88.085117}}.

\bibitem{vanSetten18}
M.~{van Setten}, M.~Giantomassi, E.~Bousquet, M.~Verstraete, D.~Hamann, X.~Gonze, G.-M. Rignanese, The pseudodojo: Training and grading a 85 element optimized norm-conserving pseudopotential table, Comput. Phys. Commun. 226 (2018) 39--54.
\newblock \href {https://doi.org/https://doi.org/10.1016/j.cpc.2018.01.012} {\path{doi:https://doi.org/10.1016/j.cpc.2018.01.012}}.

\bibitem{Methfessel89}
M.~Methfessel, A.~T. Paxton, High-precision sampling for brillouin-zone integration in metals, Phys. Rev. B 40 (1989) 3616--3621.
\newblock \href {https://doi.org/10.1103/PhysRevB.40.3616} {\path{doi:10.1103/PhysRevB.40.3616}}.

\end{thebibliography}






% \end{thebibliography}

\end{document}